\documentclass[sigconf,authorversion,pbalance]{acmart}

\AtBeginDocument{%
  }

\copyrightyear{2025}
\acmYear{2025}
\setcopyright{rightsretained}
\acmConference[SIGCSE TS 2025]{Proceedings of the 56th ACM Technical Symposium on Computer Science Education V. 1}{February 26-March 1, 2025}{Pittsburgh, PA, USA}
\acmBooktitle{Proceedings of the 56th ACM Technical Symposium on Computer Science Education V. 1 (SIGCSE TS 2025), February 26-March 1, 2025, Pittsburgh, PA, USA}
\acmDOI{10.1145/3641554.3701935}
\acmISBN{979-8-4007-0531-1/25/02}

\makeatletter
\gdef\@copyrightpermission{
  \begin{minipage}{0.3\columnwidth}
   \href{https://creativecommons.org/licenses/by/4.0/}{\includegraphics[width=0.90\textwidth]{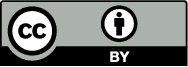}}
  \end{minipage}\hfill
  \begin{minipage}{0.7\columnwidth}
   \href{https://creativecommons.org/licenses/by/4.0/}{This work is licensed under a Creative Commons Attribution International 4.0 License.}
  \end{minipage}
  \vspace{5pt}
}
\makeatother

\begin{document}

\title{AI Technicians: Developing Rapid Occupational Training Methods for a Competitive AI Workforce}

\author{Jaromir Savelka}
\email{jsavelka@cs.cmu.edu}
\orcid{1234-5678-9012}
\affiliation{%
  \institution{Carnegie Mellon University}
  \city{Pittsburgh}
  \state{PA}
  \country{USA}
}

\author{Can Kultur}
\email{ckultur@cs.cmu.edu}
\orcid{1234-5678-9012}
\affiliation{%
  \institution{Carnegie Mellon University}
  \city{Pittsburgh}
  \state{PA}
  \country{USA}
}

\author{Arav Agarwal}
\email{arava@andrew.cmu.edu}
\orcid{1234-5678-9012}
\affiliation{%
  \institution{Carnegie Mellon University}
  \city{Pittsburgh}
  \state{PA}
  \country{USA}
}

\author{Christopher Bogart}
\email{cbogart@cs.cmu.edu}
\orcid{1234-5678-9012}
\affiliation{%
  \institution{Carnegie Mellon University}
  \city{Pittsburgh}
  \state{PA}
  \country{USA}
}

\author{Heather Burte}
\email{hburte@andrew.cmu.edu}
\orcid{1234-5678-9012}
\affiliation{%
  \institution{Carnegie Mellon University}
  \city{Pittsburgh}
  \state{PA}
  \country{USA}
}

\author{Adam Zhang}
\email{yufanz@andrew.cmu.edu}
\orcid{1234-5678-9012}
\affiliation{%
  \institution{Carnegie Mellon University}
  \city{Pittsburgh}
  \state{PA}
  \country{USA}
}

\author{Majd Sakr}
\email{msakr@andrew.cmu.edu}
\orcid{1234-5678-9012}
\affiliation{%
  \institution{Carnegie Mellon University}
  \city{Pittsburgh}
  \state{PA}
  \country{USA}
}

\renewcommand{\shortauthors}{Jaromir Savelka et al.}

\begin{abstract}
The accelerating pace of developments in Artificial Intelligence~(AI) and the increasing role that technology plays in society necessitates substantial changes in the structure of the workforce. Besides scientists and engineers, there is a need for a very large workforce of competent AI technicians (i.e., maintainers, integrators) and users~(i.e., operators). As traditional 4-year and 2-year degree-based education cannot fill this quickly opening gap, alternative training methods have to be developed. We present the results of the first four years of the AI Technicians program which is a unique collaboration between the U.S. Army's Artificial Intelligence Integration Center (AI2C) and Carnegie Mellon University to design, implement and evaluate novel rapid occupational training methods to create a competitive AI workforce at the technicians level. Through this multi-year effort we have already trained 59 AI Technicians. A key observation is that ongoing frequent updates to the training are necessary as the adoption of AI in the U.S. Army and within the society at large is evolving rapidly. A tight collaboration among the stakeholders from the army and the university is essential for successful development and maintenance of the training for the evolving role. Our findings can be leveraged by large organizations that face the challenge of developing a competent AI workforce as well as educators and researchers engaged in solving the challenge.
\end{abstract}




\maketitle

\section{Introduction}
\begin{figure}
    \centering
    \includegraphics[width=0.47\textwidth]{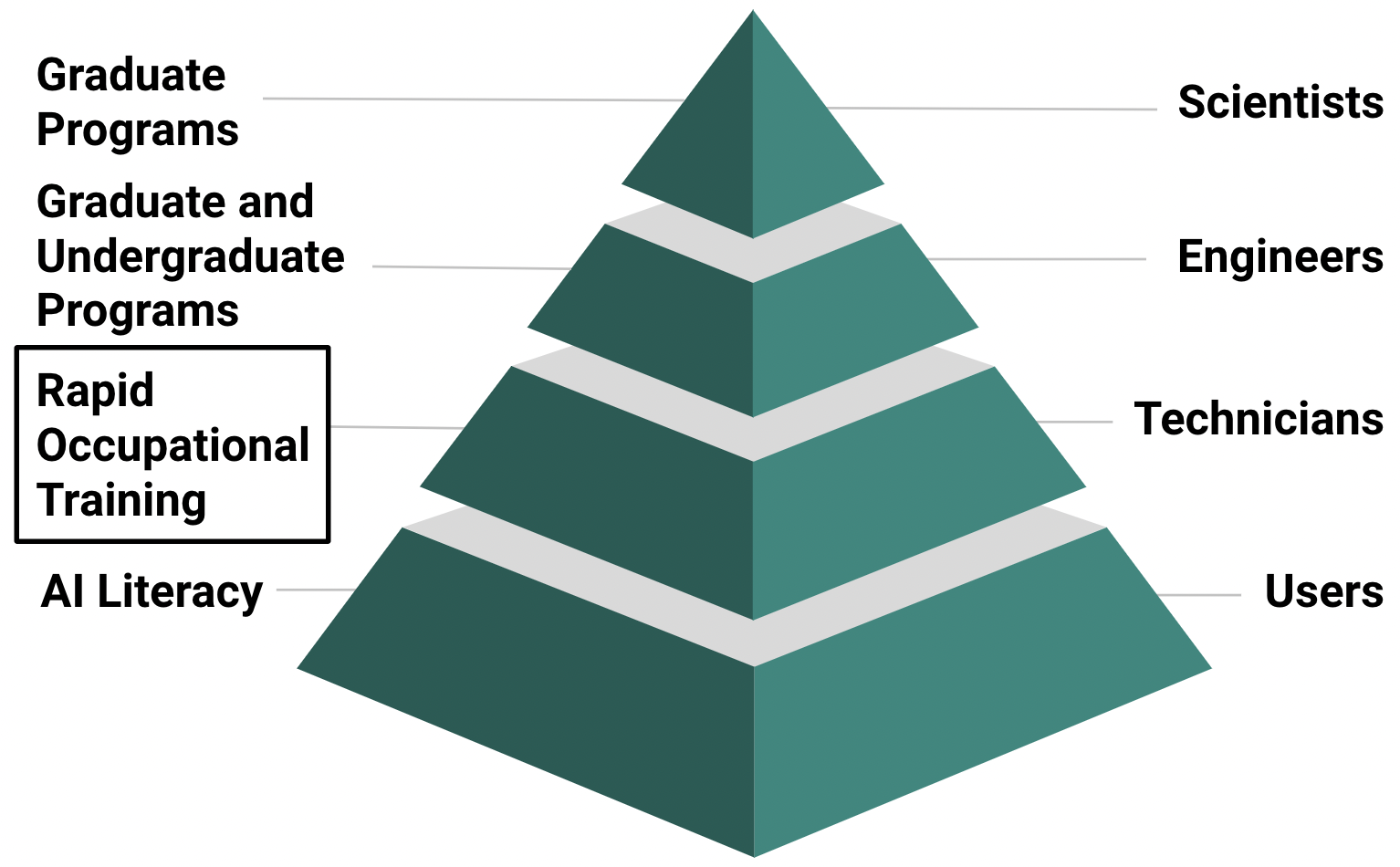}
    \caption{Rapid Occupational Training within the AI workforce structure that is illustrating different skill sets needed at each level.}
    \label{fig:ai-education}
\end{figure}

The rapid evolution and adoption of Artificial Intelligence (AI) is reshaping the way work is done in many industries, demanding a transformation in workforce structure. Beyond the scientists and engineers who innovate and advance AI capabilities, there is a growing need for AI technicians---professionals who support and maintain AI infrastructure, integrate AI solutions, and ensure seamless operation in diverse environments (see Figure \ref{fig:ai-education} illustrating the structure of the AI workforce and their training/education). The emerging role of AI Technicians has yet to be precisely defined which presents a challenge in developing effective training programs. 

The AI Technicians program is a pioneering initiative aimed at addressing this critical gap. This project is a collaborative effort between the U.S. Army's Artificial Intelligence Integration Center (AI2C) and Carnegie Mellon University, to design, implement, and evaluate novel rapid occupational training methods. The goal is to create a competitive AI workforce capable of supporting the dynamic and rapidly evolving AI-related needs of a large organization such as the U.S. Army.

The role of AI Technicians is to enable and accelerate the development, integration, and adoption of effective and timely AI-powered solutions within the U.S. Army. Their work involves building innovative AI applications and data streams, advancing AI workforce development, transforming platforms, establishing AI governance and partnerships, contributing to the AI literacy within the U.S. Army, and promoting ethical use of AI. Located near CMU's Pittsburgh campus, AI2C leverages the university's AI and robotics communities for its pioneering research work, and has been collaborating with the Technology for Effective and Efficient Learning (TEEL) Lab to build a robust training program for AI Technicians.
Developing the training program for a role that was initially undefined and that has been constantly evolving poses unique set of challenges. As AI2C grows and evolves its mission, the roles and responsibilities within its AI workforce are also evolving. 
Hence, the training program must be flexible and responsive to the changing landscape of AI technologies, their applications, and the AI2C's and the U.S. Army's needs.

The AI Technicians program, delivered in-person, has trained 59 individuals over the past four years, adapting to the rapid changes in AI technologies, and to an improved understanding of what roles can be filled by rapidly-trained and on-boarded newcomers. Frequent updates to the training content and procedures are thus necessary to keep pace with technological advancements and changing organizational needs. A tight collaboration among stakeholders from AI2C and TEEL Lab has been essential for the successful development, operation and evaluation of the training program, ensuring that it remains relevant and effective.

In this paper, we present an overview of the AI Technicians program, highlighting the iterative development of the training curriculum, the outcomes achieved, and the lessons learned. Our findings provide valuable insights to other organizations, educational institutions, and individual educators involved in developing a rapidly-trained competent workforce in an emerging domain such as AI.

\section{Related Work}
In emerging technology domains, there is often a deliberate effort to study and systematize the skill set that professionals have, in order to define targets for training; for example in cybersecurity~\cite{haney_cybersecurity_2021}. 
In other fields, such as management \cite{adriansen_two_2013}, the approach has been to explicitly teach flexibility and reflective practice, so that broadly-trained individuals can adapt themselves to the varied workplace situations they encounter. In our context, the goals were different: minimizing training time to productivity in an engineering and research facility (AI2C), and operating in a domain somewhat different than broader industry (military applications of AI). Hence, an iterative, co-evolving and co-design approach was appropriate.

Cohort learning refers to an educational approach in which a group of students progress through a program or course of study together, often following a set curriculum and timeline. This approach emphasizes the formation of a cohesive learning community, where students collaborate, support each other, and share educational experiences. Scribner and Donalson explored group dynamics and the complexity of cohort learning~\cite{paredes_scribner_dynamics_2001}. They suggest that cohort learning can prioritize affective learning over other kinds of learning because of the intense social relations existing in the cohort. They also point out the cognitive trap of thinking of learning and performance as synonyms. The cohort learning model offers several potential benefits for students such as the sense of community and belonging, promotion of peer learning and information sharing, exposure to diverse perspectives of peers, and a sense of progression and accountability as they move forward as a cohort with enhanced engagement, motivation, and satisfaction. For example, the increased sense of belonging has been recognized as especially important at 2-year colleges, where many students are already in the workforce trying to learn new skills and build confidence that they belong in their new roles  \cite{gopalan2020college}. Overall, cohort learning offers a holistic and community-oriented approach to education. This can positively impact students' success by fostering a sense of community, collaboration, and support, ultimately contributing to a more enriching and fulfilling educational experience.

Cohort-based learning is of course not new in military settings; for example Mack et al.\cite{mack_midshipmen_2019} descsribe a cohort-learning based training, for a mix of STEM and non-STEM NROTC (Naval Reserve Officer Training Corps) students in a 15-week basic Python programming skills, and a group programming project to build a cybersecurity tool. While this training was successful at giving trainees a general interaction to cybersecurity, it did not train them for a particular role, such as the AI Technicians program intends to do. 

AI's rapid advancement raises concerns about job displacement and potential transformation of human job roles \cite{acemoglu2020robots,agrawal2019artificial}. Many educational institutions and organizations are recognizing the importance of incorporating AI into their graduate and undergraduate programs. Most such educational and training efforts on AI focus on STEM students and professionals (i.e., the top two levels of the pyramid shown in Figure \ref{fig:ai-education}). Hence, there is a notable emphasis on topics such as statistics, machine learning algorithms, or formal evaluation protocols. More approachable materials for learners aspiring to become technicians or operators of AI technologies (i.e., the bottom two levels of the pyramid shown in Figure \ref{fig:ai-education}) are much less common. Yet, there is a great need for such materials \cite{ng2021conceptualizing}.

AI literacy has been emerging as a new skill set that everyone should learn, similar to classic literacy which includes reading/writing and mathematical abilities \cite{ng2021conceptualizing}. More broadly, the emergence of the knowledge-based society requires every citizen to be digitally literate, i.e., to possess basic competencies in working with digital media, computers, and more recently AI \cite{kong2021evaluation}. Specifically, AI literacy can be defined as the set of competencies crucial for effectively understanding, interacting with, and utilizing AI technologies in numerous contexts. This literacy involves more than basic awareness. It includes critical evaluation of AI technologies, effective communication and collaboration with AI systems, and their practical application in diverse settings \cite{long2020ai}. Students need to learn how to use AI judiciously, as well as to discriminate between ethical and unethical practices \cite{rodriguez2020introducing}. AI literacy means having the essential abilities that people need to live, learn and work in our digital world through AI-driven technologies, and these should be continuously honed at the K-12 levels \cite{steinbauer2021differentiated} as well as in higher education \cite{laupichler2022artificial}. For example, the ``Elements of AI'' course created by MinnaLearn and the University of Helsinki in 2018 aims to ``demystify'' AI for the public, and teach a broad audience about what AI is, how it is created, how it affects people's lives, and its likely future developments \cite{bouri2021elements}. There are also examples of domain specific materials for, e.g., lawyers \cite{savelka2020law} or radiologists \cite{allen2019democratizing}.

At the technicians level (i.e., maintainers, integrators) there is a notable lack of common approach. While formal graduate or undergraduate programs appear to go far beyond what is needed, the general AI literacy education does not provide the necessary skills and knowledge needed to perform at this level. Most large organizations develop their custom on-the-job training with varying levels of success \cite{lee2022study}. Here, we describe the rapid occupational training as a viable solution to address the workforce needs at this level.

\section{AI Technicians Program}
The AI Technicians program is a multi-year cooperation between the U.S. Army's AI2C and CMU the goal of which is to develop and evaluate novel effective methods for rapid occupational training in emerging domains, AI in particular. In this section, we describe the instrumentation of the research activities, the curricular development efforts, as well as the AI Technicians training in its current state.

\subsection{Research Instrumentation}
\label{sec:research_instrumentation}
We have used extensive instrumentation to build a rich qualitative and quantitative view of trainees' behaviors, attitudes, and outcomes resulting from the training cycle shown in Figure \ref{fig:ai2c-cycle}. The most notable instruments are the following:

\begin{figure}
    \centering
    \includegraphics[width=0.44\textwidth]{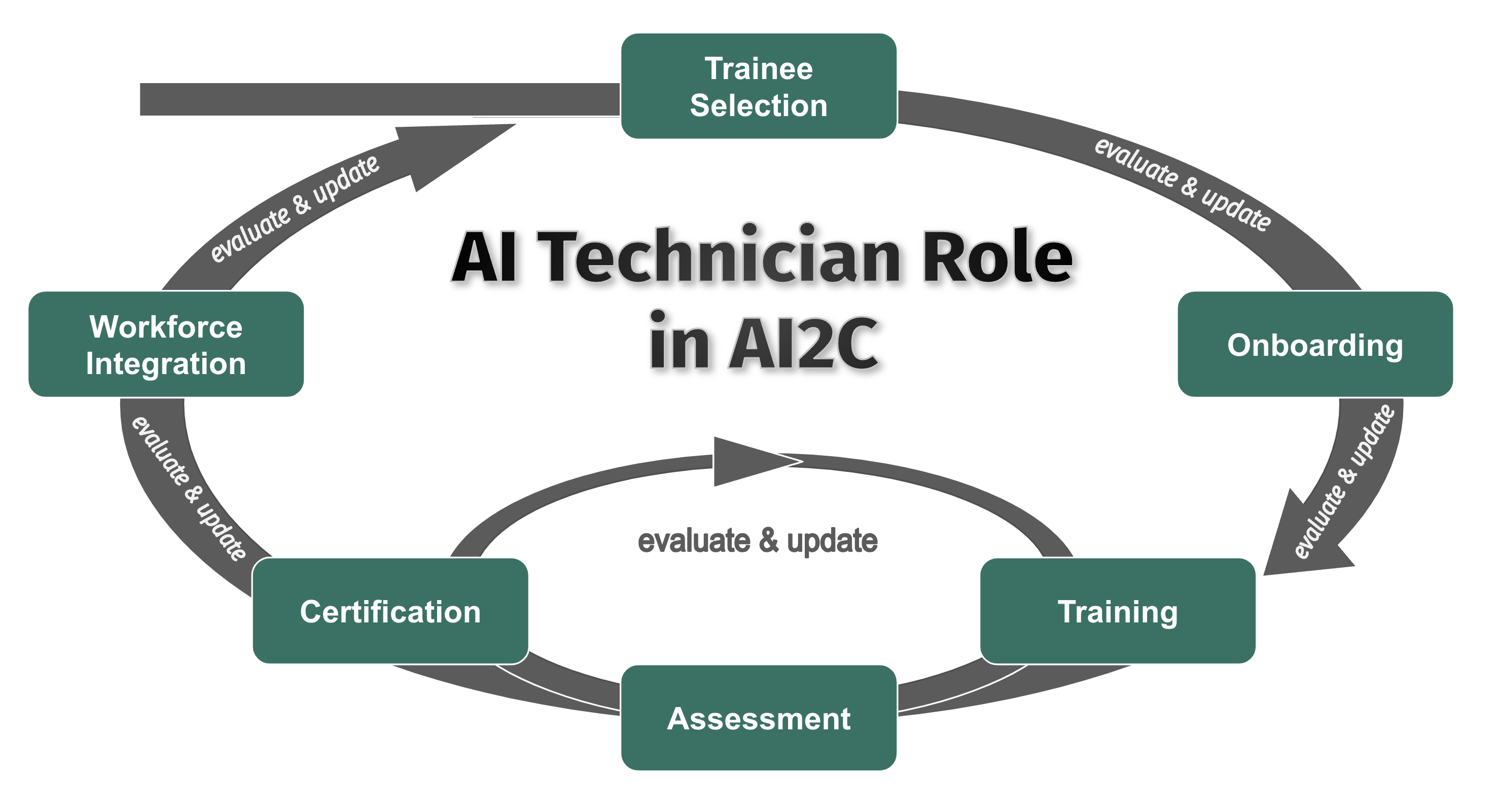}
    \caption{The training cycle of the AI technicians. The research instrumentation to measure the success of the training as well as to inform curricular development efforts is an integral part of the cycle.}
    \label{fig:ai2c-cycle}
\end{figure}

\begin{figure*}
    \centering
    \includegraphics[width=0.98\textwidth]{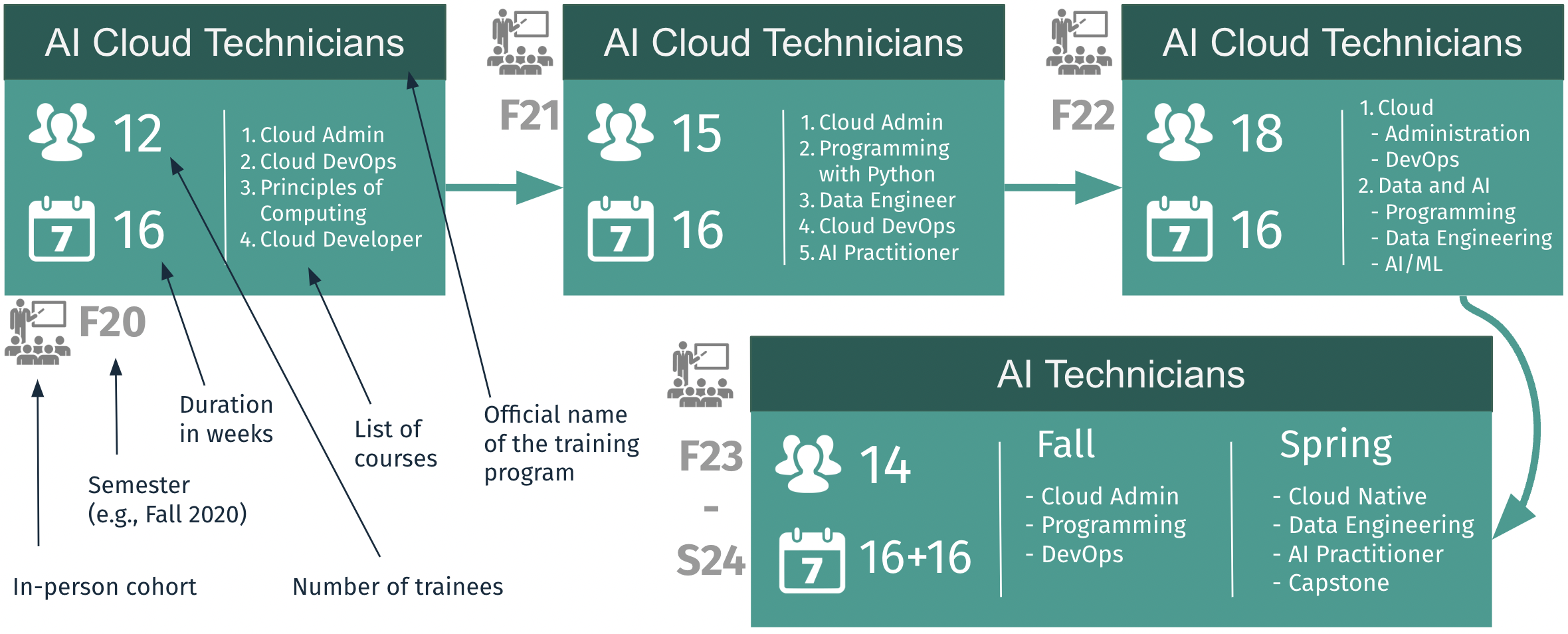}
    \caption{The evolution of the training curriculum reflects the evolution of the AI Technician role between 2020 and 2024. The original 16 weeks (one semester) long training has been extended to 32 weeks (2 semesters).}
    \label{fig:curriculum}
\end{figure*}

\subsubsection{Surveys}
We administered a suite of surveys, some only at the start of the term, and some repeated at the end, to capture aspects of trainees' background, plans, attitudes, and aptitudes, and when appropriate, changes in those attributes:
\begin{itemize}
    \item \textbf{A background survey} asked eleven questions about trainee characteristics such as ethnicity, gender, age, working hours and similar questions.
\item \textbf{A frame of mind survey} asked 28 questions adapted from published self-efficacy~\cite{Steinhorst2020}, belongingness~\cite{Knekta2020}, and STEM-identity~\cite{McDonald2019} questionnaires. The survey used a five-point Likert scale. 
In this paper we focus on self-efficacy which is trainee's  confidence in their ability to accomplish specific tasks targeted by learning objectives in the training.
\item \textbf{A knowledge and skills test}, administered at the start and the end of each semester, helped the instructor know the baseline knowledge trainees were starting from, as well as gauge effectiveness of the course. The questions were designed to test the training's learning objectives.
\item \textbf{Attitudes to AI survey} was an adaptation of existing instruments including Edison and Geissler’s ``affinity for technology'' scale \cite{edison_measuring_2003} to the domain of AI.
\end{itemize}

\subsubsection{Logging}
Our own Sail() platform (see Section \ref{sec:learning_platform}) gathers learning analytics for research, including trainee navigation and presence on all content and project pages in the platform; the timing, content, and score of each assignment submission, final project scores and a breakdown by rubric item, and content and timing of required participation in an online reflection and discussion forum after each assignment. 


\subsubsection{Qualitative instrumentation} We employ the following qualitative research instruments:

\begin{itemize}
    \item \textbf{Peer evaluations} Students working in groups on the capstone project were periodically asked to rate each other's participation.
    \item \textbf{Evaluations from stakeholders} Milestone presentations throughout the capstone process were evaluated by CMU instructors as well as AI2C project staff. 
    \item \textbf{Trainees focus groups} performed at the end of the training allowed collection of data on trainees' perspectives on training difficulty and appropriateness.  When feasible, we ran these immediately after the training, and a few months after their on-boarding in their new jobs to determine if their views had changed.
    \item \textbf{Supervisors focus groups} after each cohort completed the training and had settled into their new jobs. The sessions were focused on investigating the fit between the training and the on-the-job needs as they played out in practice.
    \item \textbf{Leadership focus groups} allowed us to gather information about the longer-term plans of the AI2C and its vision for the role of AI Technicians.
\end{itemize}

\subsection{Curriculum Development}
Originally (in 2020), the role of an AI Technician (at the time titled as ``AI Cloud Technician'') was loosely defined as to enable AI solutions by provisioning and maintaining the necessary infrastructure in public, private, edge or hybrid cloud as well as on premise, and to deploy and monitor AI-powered solutions. Over the subsequent years the role has been continuously re-defined to include awareness and understanding of emerging relevant problems, data manipulation and transformation, as well as supporting the development of the AI-powered solutions. The evolution of the role is reflected in the evolution of the training curriculum between 2020 and 2024 as shown in Figure \ref{fig:curriculum}. Notably the original one semester training (16 weeks) was eventually extended to two semesters (32 weeks). The following courses have been included in the training:

\paragraph{Cloud Administrator} This course teaches trainees to provision, orchestrate, scale, manage and monitor cloud services across compute, storage, networking, and security using various cloud interfaces. All projects utilize existing public cloud infrastructure, tools, and services. 

\paragraph{Cloud DevOps} Cloud DevOps combines people, processes, and technologies in order to increase software delivery velocity and improve service reliability. Trainees design and implement strategies for application and infrastructure that enable continuous integration, testing, continuous delivery, infrastructure as code as well as monitoring. 

\paragraph{Practical Programming with Python} Trainees learn the concepts, techniques, skills, and tools needed for developing programs in Python. Core topics include types, variables, functions, iteration, conditionals, data structures, classes, objects, modules, and I/O operations.

\paragraph{Data Engineer} The topics include ingesting, egressing and transforming data from multiple sources using various technologies, services and tools. Trainees develop the skills needed to identify and meet data requirements of an organization by designing and implementing systems and data pipelines that manage, monitor and secure the data using the full stack of cloud services. 

\paragraph{AI Practitioner} AI Practitioner provides a thorough overview of AI and machine learning (ML) applications, with a strong emphasis on hands-on experience in developing and integrating AI/ML capabilities into software solutions, deploying and maintaining AI/ML components, and evaluating AI/ML-powered systems.

\paragraph{Cloud Native} The course focuses on cloud service models and providers, microservices and architecture evaluation, and effective deployment of databases in the cloud. Trainees
develop the skills needed to handle inter-service communication, create ETL workflows, build CI pipelines, operate various cloud-native development tools, containerization technologies, and serverless computing frameworks.

\subsection{Training}

\subsubsection{Participants}
Participants are all adult learners who have been employees of the U.S. Army. For the first iteration of the training the trainees have been chosen somewhat randomly to represent the diverse workforce available across the organization. In the subsequent iterations, the AI2C has been focusing the selection criteria towards improving success of the trainees in the training program and especially in their subsequent job placement as AI Technicians at AI2C. The education levels of trainees range from a high school education, to those with completed graduate degrees.

\subsubsection{Learning Platform}
\label{sec:learning_platform}
The Sail() platform \cite{an2021working,bogart2024factors} uses a project-based learning (PBL) model. Each course begins with a set of learning objectives and incorporates a scenario that uses real-world data, tasks, and infrastructure. The courses integrate (i) learning, reflection, and feedback cycles; (ii) role-based group learning activities; and (iii) rubric-driven peer code-review opportunities. Sail() courses teach the skills for specific industry roles that can be verified by vendor-specific and vendor-neutral industry certifications. The courses are often titled with a professional role, indicating the career for which the trainee can expect to develop knowledge, skills, and certification preparation. Sail() provides ongoing formative assessment and automated contextualized feedback about assignments in progress. It includes a faculty dashboard that reports on learner activity and an anonymized data pipeline that enables research into teaching and learning. 

\subsubsection{Training Activities} The central learning activity in the AI Technicians program is a set of project-based learning courses. The hands-on projects are designed to reflect real-world processes, constraints, tasks, and industry-standard quality measures \cite{pblworks2019}. By simulating professional practice, we expose trainees to industry norms such as unit testing, code refactoring, and troubleshooting, alongside the nominal learning objectives of each project \cite{chen_pbl_impact_2015}. Our project-based learning approach encourages iterative problem-solving~\cite{pblworks2019}. To guide learners through these problems, we use several scaffolding techniques: primers~\cite[p.106]{ambrose2010learning} that scaffold prerequisite skills as worked examples, starter code that allows trainees to focus on just the relevant part of the project~\cite{quintana2004scaffolding}, and auto-grader feedback that may be requested repeatedly, to give information about trainees' partial solutions that helps them continue iterating on their work. The intent is to foster self-directed learning through actionable advice about trainees' work. In order to prompt consolidation of learning and promote social learning, after each project is completed, trainees are asked to write a reflection on their process in solving the project, and comment on each other's reflections.

\subsubsection{Capstone Projects}
In the last year, the program was extended to a second semester, in part to include a capstone experience. The capstone projects in the AI Technicians program are designed to enculturate trainees into the AI2C environment, familiarizing them with ongoing projects, work culture, and key personnel. Over a structured, four-phase process lasting sixteen weeks, trainees collaborate with university mentors, organizational supervisors, university faculty, and teaching assistants. Each phase---covering requirements gathering and design, data engineering and model development, deployment and integration, and evaluation and redeployment---includes weekly meetings, mid-phase and comprehensive end-phase presentations. This highly iterative approach with many opportunities for feedback is designed to allow trainees not only to apply their acquired knowledge but also begin the process of integrating into the organization’s operations.

\begin{figure}
    \centering
    \includegraphics[width=0.47\textwidth]{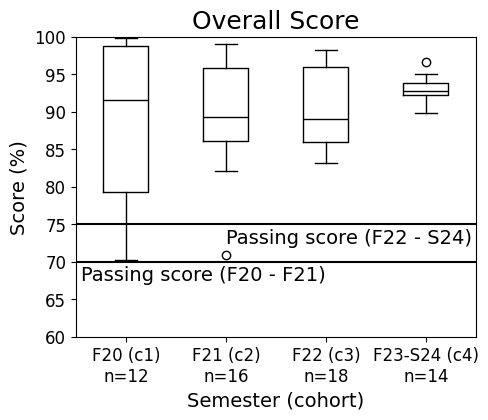}
    \caption{Distribution of overall scores from the four iterations of the training. The content of the training was changing which makes the iterations not directly comparable. The decreasing variance in scores suggests improvements in the curriculum and teaching methods as well as better targeted selection of the trainees.}
    \label{fig:overall_score}
\end{figure}

\begin{figure*}
    \centering
    \includegraphics[height=0.372\textheight]{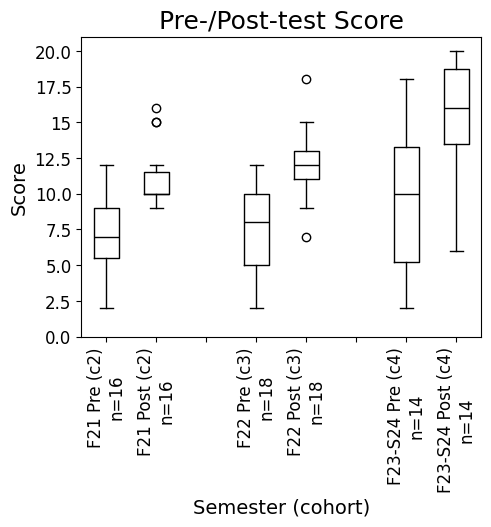}
    \includegraphics[height=0.372\textheight]{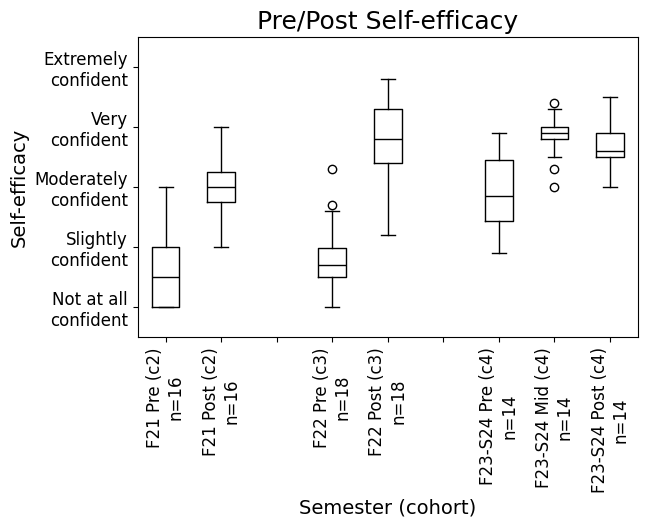}
    \caption{Left: Pre/post knowledge check scores distribution over the last three iterations of the training. Right: Pre/post self-efficacy scores over the last three iterations of the training.}
    \label{fig:pre-post}
\end{figure*}

\section{Results and Discussion}


In this section, we report selected results (mostly quantitative) acquired through the research instrumentation as described in Section \ref{sec:research_instrumentation}. As this paper is meant as an initial overview of the AI Technicians program and the surrounding efforts we do not provide a detailed presentation of all the results, especially those acquired through Sail() logging and qualitative instruments. We will continue to present more fine-grained results in the subsequent publications.

The AI Technicians program has shown consistent improvements over the four years in terms of trainee performance, knowledge acquisition, and self-efficacy. Figure~\ref{fig:overall_score} shows the distribution of scores across the four iterations. Note that the content as well as the difficulty level have been increasing over the iterations. Hence, the iterations are not directly comparable. More importantly, the variance in scores for the initial iteration was substantial, with scores ranging from slightly above the then passing mark of 70 to a perfect score of 100. However, as the development of the training program progressed, the variance narrowed, and in the last iteration, not only did the scores reach their highest median, but the distribution was also the narrowest. This improvement in consistency can be attributed to refinements in the curriculum and teaching methods as well as to better targeted selection of the trainees. 

Figure~\ref{fig:pre-post} presents the pre- and post-term knowledge assessments and self-efficacy measurements over the past three iterations of the training. The knowledge check scores (a) have shown increases from the start to the end of each term, with both pre- and post-term scores rising each year. The last year stands out with the highest scores. The self-efficacy scores (Figure~\ref{fig:pre-post}) increased notably from the start to the end of each term. Similar to the knowledge check scores, the self-efficacy scores have started higher each year, reflecting the increasing confidence of trainees in their abilities (likely due to better targeted selection of the trainees). However, in the last iteration, the increase in self-efficacy was less pronounced, possibly due to the high demands of concurrently completing the capstone project and the challenging Cloud Native course.

Over the four-year period, the iterative development of the AI Technicians program as well as the changes in trainee selection have led to substantial improvements in both the performance and self-efficacy of trainees. We hypothesize that one of the key strengths of the program enabling this kind of success is its adoption of PBL. By engaging trainees in realistic projects that replicate real-world industry scenarios, the program ensures that participants gain valuable hands-on experience and build self-efficacy. By going through the training as a cohort, trainees also benefit from peer support and collaborative learning. We hypothesize that this helps to build a sense of community and shared purpose which in turn enhances their motivation and engagement.


\paragraph{Implications for Teaching Practice}
Some of our lessons learned from developing the AI Technicians may be relevant to teaching practice in other rapidly evolving technological fields. An adaptive curriculum, continuously updated based on stakeholder feedback, ensures that the training remains relevant and effective. Incorporating PBL allows learners to engage with real-world scenarios and develop practical, hands-on skills that are directly applicable to their future roles. The cohort-based model enhances learning by fostering a sense of community and peer support, which can significantly boost engagement and retention.

\paragraph{Limitations}
Some of the conditions that lead to the success of the AI Technicians program may be specific to our context, and difficult to replicate in less ideal circumstances.
The AI Technicians program has 15-25 learners per annual cohort.
The program itself has also evolved over the years, with major additions and minor removals of content during each iteration, tracking evolving understanding of AI2C's needs, but making fair comparisons across years more difficult. We also need to emphasize that the trainee selection process has evolved over time to target more technically apt individuals and continues to evolve in terms of increasing diversity to support trainees with various backgrounds. Finally, different from many common training program settings (at community colleges or on the job), the AI Technicians trainees are enrolled full-time, in-person, and taking all courses together at the same time as a cohort. This gives them more opportunities to interact with each other and support each other. These favorable conditions are a cost that other organizations may not be able to bear, but we believe this model can be applied when it becomes critical to perform rapid training in a rapidly-evolving field.

\section{Conclusions and Future Work}
The AI Technicians program has successfully demonstrated the viability of rapid occupational training methods tailored to the dynamic needs of the AI workforce. Through an iterative, adaptive curriculum co-designed with the stakeholders which focuses on PBL and cohort-based learning, the program has effectively prepared trainees for roles that require both practical skills and the ability to adapt to continuous technological advancements in AI. Future work should focus on scaling the AI Technicians training model to accommodate larger cohorts and different organizational contexts within the US Army. This may also include exploring hybrid and fully online delivery methods to increase accessibility and reduce costs. Longitudinal studies tracking the career progression and on-the-job performance of program graduates will provide deeper insights into the long-term impact of the training. 

\begin{acks}
This material is based upon work supported by the U.S. Army Research Office and the U.S. Army Futures Command under Contract No. W519TC-23-C-0043. The content of the information does not necessarily reflect the position or the policy of the government and no official endorsement should be inferred.
\end{acks}

\bibliographystyle{ACM-Reference-Format}
\bibliography{sample-base,from_zotero}


\begin{thebibliography}{27}


\ifx \showCODEN    \undefined \def \showCODEN     #1{\unskip}     \fi
\ifx \showDOI      \undefined \def \showDOI       #1{#1}\fi
\ifx \showISBNx    \undefined \def \showISBNx     #1{\unskip}     \fi
\ifx \showISBNxiii \undefined \def \showISBNxiii  #1{\unskip}     \fi
\ifx \showISSN     \undefined \def \showISSN      #1{\unskip}     \fi
\ifx \showLCCN     \undefined \def \showLCCN      #1{\unskip}     \fi
\ifx \shownote     \undefined \def \shownote      #1{#1}          \fi
\ifx \showarticletitle \undefined \def \showarticletitle #1{#1}   \fi
\ifx \showURL      \undefined \def \showURL       {\relax}        \fi
\providecommand\bibfield[2]{#2}
\providecommand\bibinfo[2]{#2}
\providecommand\natexlab[1]{#1}
\providecommand\showeprint[2][]{arXiv:#2}

\bibitem[Acemoglu and Restrepo(2020)]%
        {acemoglu2020robots}
\bibfield{author}{\bibinfo{person}{Daron Acemoglu} {and} \bibinfo{person}{Pascual Restrepo}.} \bibinfo{year}{2020}\natexlab{}.
\newblock \showarticletitle{Robots and jobs: Evidence from US labor markets}.
\newblock \bibinfo{journal}{\emph{Journal of political economy}} \bibinfo{volume}{128}, \bibinfo{number}{6} (\bibinfo{year}{2020}), \bibinfo{pages}{2188--2244}.
\newblock


\bibitem[Adriansen and Knudsen(2013)]%
        {adriansen_two_2013}
\bibfield{author}{\bibinfo{person}{Hanne~Kirstine Adriansen} {and} \bibinfo{person}{Hanne Knudsen}.} \bibinfo{year}{2013}\natexlab{}.
\newblock \showarticletitle{Two ways to support reflexivity: {Teaching} managers to fulfil an undefined role}.
\newblock \bibinfo{journal}{\emph{Teaching Public Administration}} \bibinfo{volume}{31}, \bibinfo{number}{1} (\bibinfo{year}{2013}).
\newblock
\urldef\tempurl%
\url{https://doi.org/10.1177/0144739412474457}
\showDOI{\tempurl}


\bibitem[Agrawal et~al\mbox{.}(2019)]%
        {agrawal2019artificial}
\bibfield{author}{\bibinfo{person}{Ajay Agrawal}, \bibinfo{person}{Joshua~S Gans}, {and} \bibinfo{person}{Avi Goldfarb}.} \bibinfo{year}{2019}\natexlab{}.
\newblock \showarticletitle{Artificial intelligence: the ambiguous labor market impact of automating prediction}.
\newblock \bibinfo{journal}{\emph{Journal of Economic Perspectives}} \bibinfo{volume}{33}, \bibinfo{number}{2} (\bibinfo{year}{2019}), \bibinfo{pages}{31--50}.
\newblock


\bibitem[Allen et~al\mbox{.}(2019)]%
        {allen2019democratizing}
\bibfield{author}{\bibinfo{person}{Bibb Allen}, \bibinfo{person}{Sheela Agarwal}, \bibinfo{person}{Jayashree Kalpathy-Cramer}, {and} \bibinfo{person}{Keith Dreyer}.} \bibinfo{year}{2019}\natexlab{}.
\newblock \showarticletitle{Democratizing ai}.
\newblock \bibinfo{journal}{\emph{Journal of the American College of Radiology}} \bibinfo{volume}{16}, \bibinfo{number}{7} (\bibinfo{year}{2019}), \bibinfo{pages}{961--963}.
\newblock


\bibitem[Ambrose et~al\mbox{.}(2010)]%
        {ambrose2010learning}
\bibfield{author}{\bibinfo{person}{Susan~A Ambrose}, \bibinfo{person}{Michael~W Bridges}, \bibinfo{person}{Michele DiPietro}, \bibinfo{person}{Marsha~C Lovett}, {and} \bibinfo{person}{Marie~K Norman}.} \bibinfo{year}{2010}\natexlab{}.
\newblock \bibinfo{booktitle}{\emph{How learning works: Seven research-based principles for smart teaching}}.
\newblock \bibinfo{publisher}{John Wiley \& Sons}.
\newblock


\bibitem[An et~al\mbox{.}(2021)]%
        {an2021working}
\bibfield{author}{\bibinfo{person}{Mingxiao An}, \bibinfo{person}{Hongyi Zhang}, \bibinfo{person}{Jaromir Savelka}, \bibinfo{person}{Shijie Zhu}, \bibinfo{person}{Chris Bogart}, {and} \bibinfo{person}{Majd Sakr}.} \bibinfo{year}{2021}\natexlab{}.
\newblock \showarticletitle{Are Working Habits Different Between Well-Performing and at-Risk Students in Online Project-Based Courses?}. In \bibinfo{booktitle}{\emph{Proceedings of the 26th ACM Conference on Innovation and Technology in Computer Science Education V. 1}}. \bibinfo{pages}{324--330}.
\newblock


\bibitem[Bogart et~al\mbox{.}(2024)]%
        {bogart2024factors}
\bibfield{author}{\bibinfo{person}{Christopher Bogart}, \bibinfo{person}{Marshall An}, \bibinfo{person}{Eric Keylor}, \bibinfo{person}{Pawanjeet Singh}, \bibinfo{person}{Jaromir Savelka}, {and} \bibinfo{person}{Majd Sakr}.} \bibinfo{year}{2024}\natexlab{}.
\newblock \showarticletitle{What Factors Influence Persistence in Project-based Programming Courses at Community Colleges?}. In \bibinfo{booktitle}{\emph{Proceedings of the 55th ACM Technical Symposium on Computer Science Education V. 1}}. \bibinfo{pages}{116--122}.
\newblock


\bibitem[Bouri and Reponen(2021)]%
        {bouri2021elements}
\bibfield{author}{\bibinfo{person}{Ioanna Bouri} {and} \bibinfo{person}{Sanna Reponen}.} \bibinfo{year}{2021}\natexlab{}.
\newblock \showarticletitle{Elements of AI: Busting AI myths on a global scale}. In \bibinfo{booktitle}{\emph{Proceedings of the 21st Koli Calling International Conference on Computing Education Research}}. \bibinfo{pages}{1--2}.
\newblock


\bibitem[Chen et~al\mbox{.}(2015)]%
        {chen_pbl_impact_2015}
\bibfield{author}{\bibinfo{person}{Pearl Chen}, \bibinfo{person}{Anthony Hernandez}, {and} \bibinfo{person}{Jane Dong}.} \bibinfo{year}{2015}\natexlab{}.
\newblock \showarticletitle{Impact of collaborative project-based learning on self-efficacy of urban minority students in engineering}.
\newblock \bibinfo{journal}{\emph{Journal of Urban Learning, Teaching, and Research}}  \bibinfo{volume}{11} (\bibinfo{year}{2015}), \bibinfo{pages}{26--39}.
\newblock


\bibitem[Edison and Geissler(2003)]%
        {edison_measuring_2003}
\bibfield{author}{\bibinfo{person}{Steve~W Edison} {and} \bibinfo{person}{Gary~L Geissler}.} \bibinfo{year}{2003}\natexlab{}.
\newblock \showarticletitle{Measuring attitudes towards general technology: {Antecedents}, hypotheses and scale development}.
\newblock \bibinfo{journal}{\emph{Journal of Targeting, Measurement and Analysis for Marketing}} \bibinfo{volume}{12}, \bibinfo{number}{2} (\bibinfo{date}{Nov.} \bibinfo{year}{2003}), \bibinfo{pages}{137--156}.
\newblock
\showISSN{1479-1862}
\urldef\tempurl%
\url{https://doi.org/10.1057/palgrave.jt.5740104}
\showDOI{\tempurl}


\bibitem[Gopalan and Brady(2020)]%
        {gopalan2020college}
\bibfield{author}{\bibinfo{person}{Maithreyi Gopalan} {and} \bibinfo{person}{Shannon~T Brady}.} \bibinfo{year}{2020}\natexlab{}.
\newblock \showarticletitle{College students’ sense of belonging: A national perspective}.
\newblock \bibinfo{journal}{\emph{Educational Researcher}} \bibinfo{volume}{49}, \bibinfo{number}{2} (\bibinfo{year}{2020}), \bibinfo{pages}{134--137}.
\newblock


\bibitem[Haney and Lutters(2021)]%
        {haney_cybersecurity_2021}
\bibfield{author}{\bibinfo{person}{Julie~M. Haney} {and} \bibinfo{person}{Wayne~G. Lutters}.} \bibinfo{year}{2021}\natexlab{}.
\newblock \showarticletitle{Cybersecurity advocates: discovering the characteristics and skills of an emergent role}.
\newblock \bibinfo{journal}{\emph{Information \&Amp; Computer Security}} \bibinfo{volume}{29}, \bibinfo{number}{3} (\bibinfo{year}{2021}).
\newblock
\urldef\tempurl%
\url{https://doi.org/10.1108/ics-08-2020-0131}
\showDOI{\tempurl}


\bibitem[Knekta et~al\mbox{.}(2020)]%
        {Knekta2020}
\bibfield{author}{\bibinfo{person}{Eva Knekta}, \bibinfo{person}{Kyriaki Chatzikyriakidou}, {and} \bibinfo{person}{Melissa McCartney}.} \bibinfo{year}{2020}\natexlab{}.
\newblock \showarticletitle{Evaluation of a questionnaire measuring university students’ sense of belonging to and involvement in a biology department}.
\newblock \bibinfo{journal}{\emph{CBE Life Sciences Education}} \bibinfo{volume}{19}, \bibinfo{number}{3} (\bibinfo{year}{2020}), \bibinfo{pages}{1--14}.
\newblock
\showISSN{19317913}
\urldef\tempurl%
\url{https://doi.org/10.1187/cbe.19-09-0166}
\showDOI{\tempurl}


\bibitem[Kong et~al\mbox{.}(2021)]%
        {kong2021evaluation}
\bibfield{author}{\bibinfo{person}{Siu-Cheung Kong}, \bibinfo{person}{William Man-Yin Cheung}, {and} \bibinfo{person}{Guo Zhang}.} \bibinfo{year}{2021}\natexlab{}.
\newblock \showarticletitle{Evaluation of an artificial intelligence literacy course for university students with diverse study backgrounds}.
\newblock \bibinfo{journal}{\emph{Computers and Education: Artificial Intelligence}}  \bibinfo{volume}{2} (\bibinfo{year}{2021}), \bibinfo{pages}{100026}.
\newblock


\bibitem[Laupichler et~al\mbox{.}(2022)]%
        {laupichler2022artificial}
\bibfield{author}{\bibinfo{person}{Matthias~Carl Laupichler}, \bibinfo{person}{Alexandra Aster}, \bibinfo{person}{Jana Schirch}, {and} \bibinfo{person}{Tobias Raupach}.} \bibinfo{year}{2022}\natexlab{}.
\newblock \showarticletitle{Artificial intelligence literacy in higher and adult education: A scoping literature review}.
\newblock \bibinfo{journal}{\emph{Computers and Education: Artificial Intelligence}}  \bibinfo{volume}{3} (\bibinfo{year}{2022}), \bibinfo{pages}{100101}.
\newblock


\bibitem[Lee et~al\mbox{.}(2022)]%
        {lee2022study}
\bibfield{author}{\bibinfo{person}{Won~Joo Lee}, \bibinfo{person}{Doohyun Kim}, \bibinfo{person}{Sang~Il Kim}, {and} \bibinfo{person}{Han~Sung Kim}.} \bibinfo{year}{2022}\natexlab{}.
\newblock \showarticletitle{A Study on the Standard AI Developer Job Training Track Based on Industry Demand}.
\newblock \bibinfo{journal}{\emph{Journal of The Korea Society of Computer and Information}} \bibinfo{volume}{27}, \bibinfo{number}{3} (\bibinfo{year}{2022}), \bibinfo{pages}{251--258}.
\newblock


\bibitem[Long and Magerko(2020)]%
        {long2020ai}
\bibfield{author}{\bibinfo{person}{Duri Long} {and} \bibinfo{person}{Brian Magerko}.} \bibinfo{year}{2020}\natexlab{}.
\newblock \showarticletitle{What is AI literacy? Competencies and design considerations}. In \bibinfo{booktitle}{\emph{Proceedings of the 2020 CHI conference on human factors in computing systems}}. \bibinfo{pages}{1--16}.
\newblock


\bibitem[Mack et~al\mbox{.}(2019)]%
        {mack_midshipmen_2019}
\bibfield{author}{\bibinfo{person}{Naja~A. Mack}, \bibinfo{person}{Kevin Womack}, \bibinfo{person}{Earl~W. Huff~Jr.}, \bibinfo{person}{Robert Cummings}, \bibinfo{person}{Negus Dowling}, {and} \bibinfo{person}{Kinnis Gosha}.} \bibinfo{year}{2019}\natexlab{}.
\newblock \showarticletitle{From {Midshipmen} to {Cyber} {Pros}: {Training} {Minority} {Naval} {Reserve} {Officer} {Training} {Corp} {Students} for {Cybersecurity}}. In \bibinfo{booktitle}{\emph{Proceedings of the 50th {ACM} {Technical} {Symposium} on {Computer} {Science} {Education}}}. \bibinfo{publisher}{ACM}, \bibinfo{address}{Minneapolis MN USA}, \bibinfo{pages}{726--730}.
\newblock
\showISBNx{978-1-4503-5890-3}
\urldef\tempurl%
\url{https://doi.org/10.1145/3287324.3287500}
\showDOI{\tempurl}


\bibitem[McDonald et~al\mbox{.}(2019)]%
        {McDonald2019}
\bibfield{author}{\bibinfo{person}{Melissa~M. McDonald}, \bibinfo{person}{Virgil Zeigler-Hill}, \bibinfo{person}{Jennifer~K. Vrabel}, {and} \bibinfo{person}{Martha Escobar}.} \bibinfo{year}{2019}\natexlab{}.
\newblock \showarticletitle{A {Single}-{Item} {Measure} for {Assessing} {STEM} {Identity}}.
\newblock \bibinfo{journal}{\emph{Frontiers in Education}} \bibinfo{volume}{4}, \bibinfo{number}{July} (\bibinfo{year}{2019}), \bibinfo{pages}{1--15}.
\newblock
\showISSN{2504284X}
\urldef\tempurl%
\url{https://doi.org/10.3389/feduc.2019.00078}
\showDOI{\tempurl}


\bibitem[Ng et~al\mbox{.}(2021)]%
        {ng2021conceptualizing}
\bibfield{author}{\bibinfo{person}{Davy Tsz~Kit Ng}, \bibinfo{person}{Jac Ka~Lok Leung}, \bibinfo{person}{Samuel Kai~Wah Chu}, {and} \bibinfo{person}{Maggie~Shen Qiao}.} \bibinfo{year}{2021}\natexlab{}.
\newblock \showarticletitle{Conceptualizing AI literacy: An exploratory review}.
\newblock \bibinfo{journal}{\emph{Computers and Education: Artificial Intelligence}}  \bibinfo{volume}{2} (\bibinfo{year}{2021}), \bibinfo{pages}{100041}.
\newblock


\bibitem[Paredes~Scribner and Donaldson(2001)]%
        {paredes_scribner_dynamics_2001}
\bibfield{author}{\bibinfo{person}{Jay Paredes~Scribner} {and} \bibinfo{person}{Joe~F. Donaldson}.} \bibinfo{year}{2001}\natexlab{}.
\newblock \showarticletitle{The {Dynamics} of {Group} {Learning} in a {Cohort}: {From} {Nonlearning} to {Transformative} {Learning}}.
\newblock \bibinfo{journal}{\emph{Educational Administration Quarterly}} \bibinfo{volume}{37}, \bibinfo{number}{5} (\bibinfo{year}{2001}).
\newblock
\urldef\tempurl%
\url{https://journals.sagepub.com/doi/10.1177/00131610121969442}
\showURL{%
\tempurl}


\bibitem[PBLWorks(2023)]%
        {pblworks2019}
\bibfield{author}{\bibinfo{person}{PBLWorks}.} \bibinfo{year}{2023}\natexlab{}.
\newblock \bibinfo{title}{Gold {Standard} {PBL}: {The} {Essential} {Project} {Design} {Elements}}.
\newblock
\newblock
\urldef\tempurl%
\url{https://my.pblworks.org/resource/document/gold_standard_pbl_essential_project_design_elements}
\showURL{%
\tempurl}


\bibitem[Quintana et~al\mbox{.}(2004)]%
        {quintana2004scaffolding}
\bibfield{author}{\bibinfo{person}{Chris Quintana}, \bibinfo{person}{Brian~J Reiser}, \bibinfo{person}{Elizabeth~A Davis}, \bibinfo{person}{Joseph Krajcik}, \bibinfo{person}{Eric Fretz}, \bibinfo{person}{Ravit~Golan Duncan}, \bibinfo{person}{Eleni Kyza}, \bibinfo{person}{Daniel Edelson}, {and} \bibinfo{person}{Elliot Soloway}.} \bibinfo{year}{2004}\natexlab{}.
\newblock \showarticletitle{A scaffolding design framework for software to support science inquiry}.
\newblock \bibinfo{journal}{\emph{The journal of the learning sciences}} \bibinfo{volume}{13}, \bibinfo{number}{3} (\bibinfo{year}{2004}), \bibinfo{pages}{337--386}.
\newblock


\bibitem[Rodr{\'\i}guez-Garc{\'\i}a et~al\mbox{.}(2020)]%
        {rodriguez2020introducing}
\bibfield{author}{\bibinfo{person}{Juan~David Rodr{\'\i}guez-Garc{\'\i}a}, \bibinfo{person}{Jes{\'u}s Moreno-Le{\'o}n}, \bibinfo{person}{Marcos Rom{\'a}n-Gonz{\'a}lez}, {and} \bibinfo{person}{Gregorio Robles}.} \bibinfo{year}{2020}\natexlab{}.
\newblock \showarticletitle{Introducing artificial intelligence fundamentals with LearningML: Artificial intelligence made easy}. In \bibinfo{booktitle}{\emph{Eighth international conference on technological ecosystems for enhancing multiculturality}}. \bibinfo{pages}{18--20}.
\newblock


\bibitem[Savelka et~al\mbox{.}(2020)]%
        {savelka2020law}
\bibfield{author}{\bibinfo{person}{Jaromir Savelka}, \bibinfo{person}{Matthias Grabmair}, {and} \bibinfo{person}{Kevin~D Ashley}.} \bibinfo{year}{2020}\natexlab{}.
\newblock \showarticletitle{A law school course in applied legal analytics and AI}.
\newblock \bibinfo{journal}{\emph{Law Context: A Socio-Legal J.}}  \bibinfo{volume}{37} (\bibinfo{year}{2020}), \bibinfo{pages}{134}.
\newblock


\bibitem[Steinbauer et~al\mbox{.}(2021)]%
        {steinbauer2021differentiated}
\bibfield{author}{\bibinfo{person}{Gerald Steinbauer}, \bibinfo{person}{Martin Kandlhofer}, \bibinfo{person}{Tara Chklovski}, \bibinfo{person}{Fredrik Heintz}, {and} \bibinfo{person}{Sven Koenig}.} \bibinfo{year}{2021}\natexlab{}.
\newblock \showarticletitle{A differentiated discussion about AI education K-12}.
\newblock \bibinfo{journal}{\emph{KI-K{\"u}nstliche Intelligenz}} \bibinfo{volume}{35}, \bibinfo{number}{2} (\bibinfo{year}{2021}), \bibinfo{pages}{131--137}.
\newblock


\bibitem[Steinhorst et~al\mbox{.}(2020)]%
        {Steinhorst2020}
\bibfield{author}{\bibinfo{person}{Phil Steinhorst}, \bibinfo{person}{Andrew Petersen}, {and} \bibinfo{person}{Jan Vahrenhold}.} \bibinfo{year}{2020}\natexlab{}.
\newblock \showarticletitle{Revisiting {Self}-{Efficacy} in {Introductory} {Programming}}.
\newblock \bibinfo{journal}{\emph{Proc. Conf. Intl. Comp. Education Research}} (\bibinfo{year}{2020}), \bibinfo{pages}{158--169}.
\newblock
\urldef\tempurl%
\url{https://doi.org/10.1145/3372782.3406281}
\showDOI{\tempurl}


\end{thebibliography}










\end{document}